\documentclass[doublecol]{epl2}
\usepackage{graphicx}
\usepackage{amsmath}
\usepackage{amssymb}

\title{Scaling of Transport Coefficients of Porous Media\\ under Compaction}

\author{Denis S.\ Goldobin}
\shortauthor{D.\ S.\ Goldobin}
\institute{Department of Mathematics, University of Leicester,
             Leicester LE1 7RH, UK\\
           Institute of Continuous Media Mechanics, UB RAS,
             Perm 614013, Russia}

\abstract{
Porous sediments in geological systems are exposed to stress by
the above-laying mass and consequent compaction, which may be
significantly nonuniform across the massif. We derive scaling laws
for the compaction of sediments of similar geological origin. With
these laws, we evaluate the dependence of the transport properties
of a fluid-saturated porous medium (permeability, effective
molecular diffusivity, hydrodynamic dispersion, electrical and
thermal conductivities) on its porosity. In particular, we
demonstrate that the assumption of a uniform geothermal gradient
is not adequate for systems with nonuniform compaction and show
the importance of the derived scaling laws for mathematical
modelling of methane hydrate deposits; these deposits are believed
to have potential for impact on global climate change and
Glacial-Interglacial cycles.
 }

\begin{document}

\maketitle

\section{Introduction}
The reconstruction of properties of grounds is an important
problem related to geological surveys for extraction of minerals
and hydrocarbons, construction of buildings, forecasting
geological hazards (seismic, erosion-related, anomalous impact on
the climate, {\it
etc.})~\cite{Sahimi-1993,Nield-Bejan-2006,Mavko-Mukerji-Dvorkin-2009}.
Dealing with this problem one faces challenges, some of which
hardly may be overcome. These challenges are related to the
impossibility of direct measurements of required parameters across
large massifs. Even making boreholes provides limited information
about the narrow vicinity of the borehole; for instance,
measurements of the electrical conductivity and the porosity of
the porous medium are generally not enough to reconstruct its
permeability, which practically can not be measured directly.
Thus, the problem actually turns into one of the recovery of
relations between different (transport) parameters of the porous
media. Generally, this problem is non-resolvable, because it
requires thorough knowledge of the composition of the massif, its
geological and seismic history, {\it etc.} Meanwhile, many recent
studies deal with systems where the massif possesses a homogenous
geological origin on the field
scale~\cite{Proc_ODP-V164-InitResults, Proc_ODP-V164,
Davie-Buffett-2001, Davie-Buffett-2003, Archer-2007,
Garg_etal-2008}. Opportunities for an advance in the problem of
reconstruction of relations between parameters for such kinds of
systems might lay in the field of mathematical physics. In this
study we wish to approach this problem in application to some
important geological systems, like ocean bed with methane hydrate
deposits.

\begin{figure}[!t]
\centerline{
 \includegraphics[width=0.23\textwidth]%
 {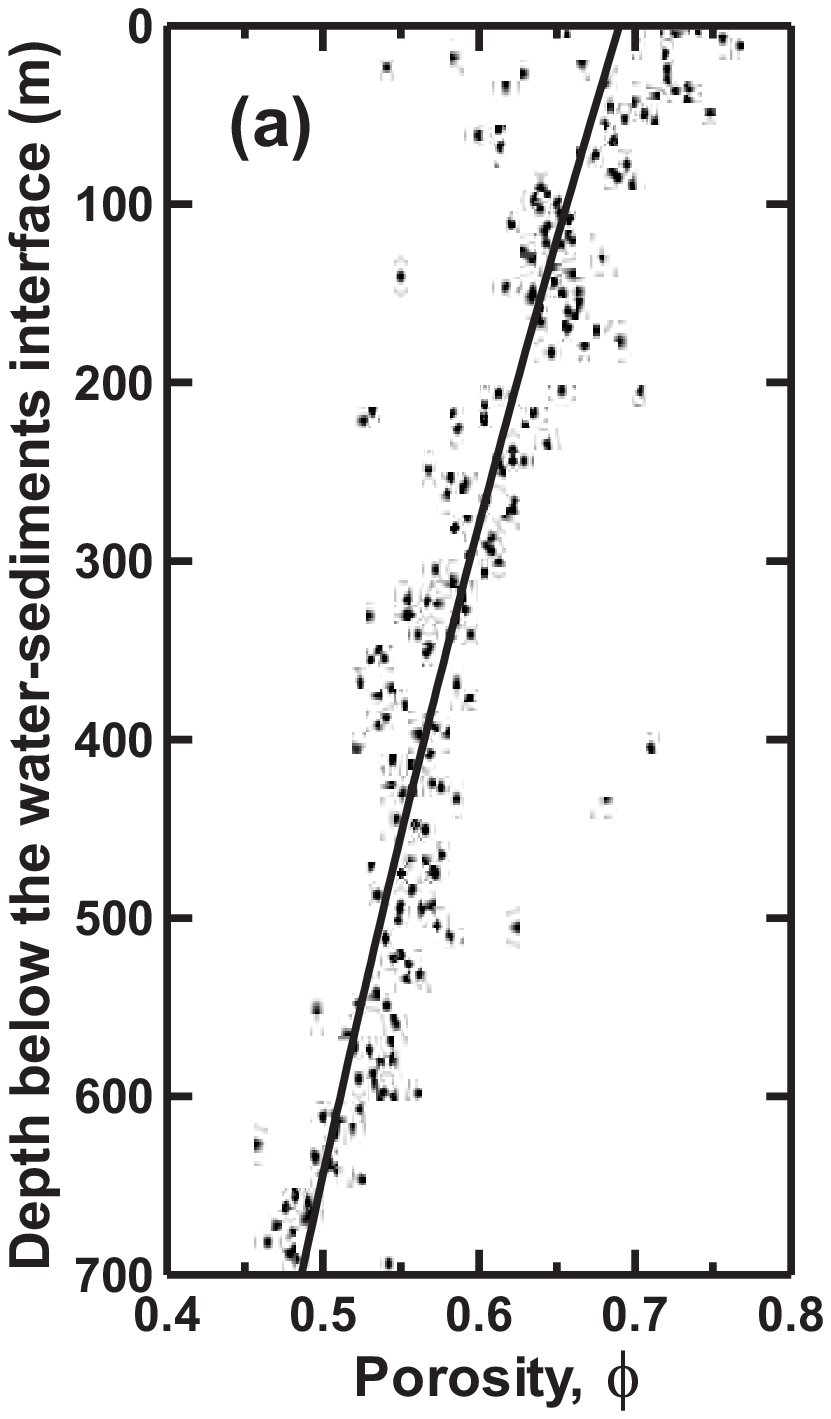}
\quad
 \includegraphics[width=0.23\textwidth]%
 {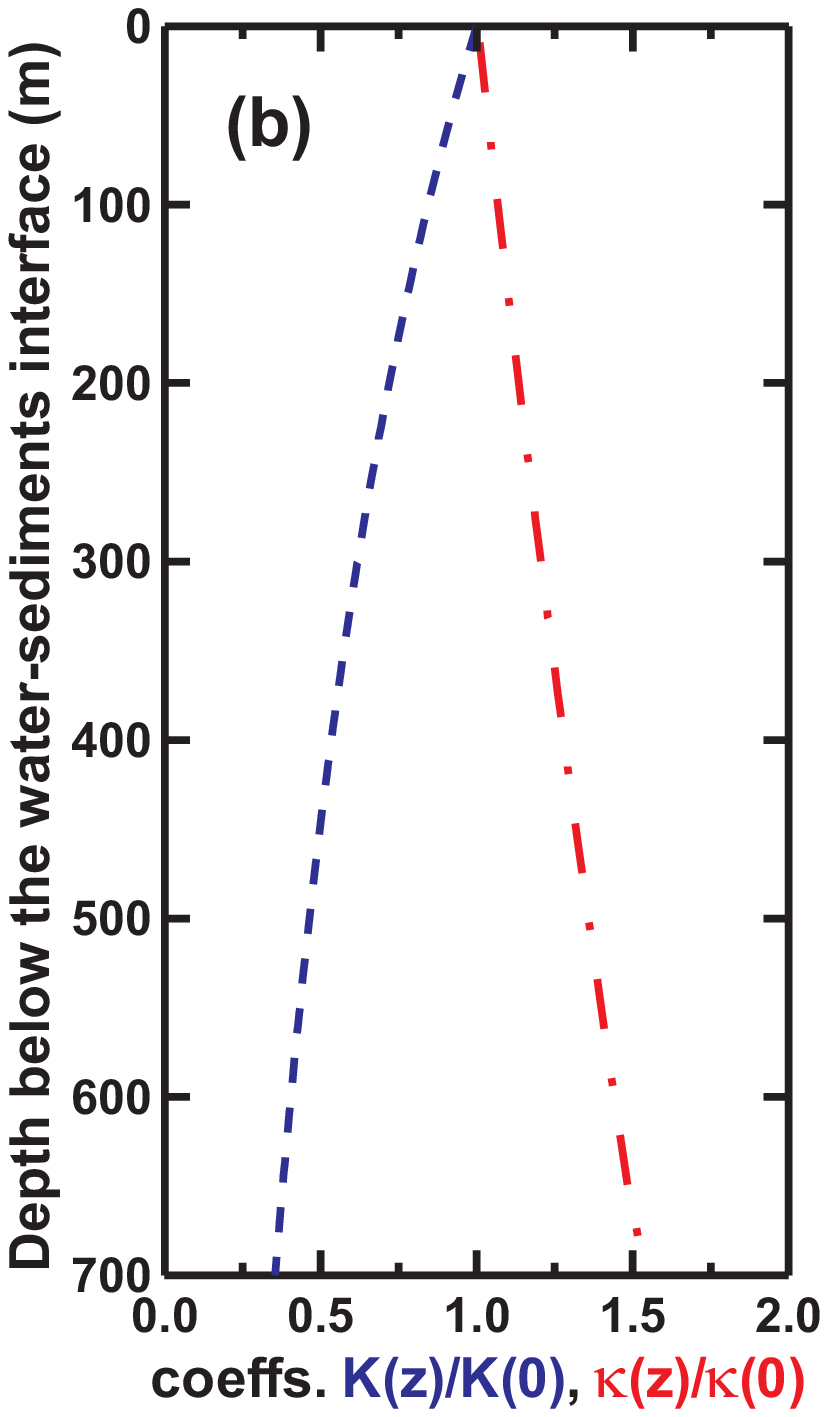}
}
  \caption{(a)~Observed porosity vs the depth below
the water-sediment interface for the ODP (Ocean Drilling
Program~\cite{Proc_ODP-V164-InitResults,Proc_ODP-V164}) site 997
on the Blake Ridge crest---one of the largest marine hydrate
provinces. The solid line represents the exponential decrease of
the porosity (Eq.\,(\ref{eq-in-01}) with $\phi_0=0.69$ and
$L=2$km~\cite{Davie-Buffett-2001}). (b)~Predicted permeability
[blue dashed line, Eq.\,(\ref{eq-an-03})] and thermal conductivity
[red dash-dot line, Eq.\,(\ref{eq-an-08})] of sediments.
 }
  \label{fig1}
\end{figure}

The ocean bed in regions with intense mud rain is very attractive
and important for research due to bio- and geological richness and
activity~\cite{Proc_ODP-V164-InitResults,Proc_ODP-V164}; for
example, these ocean bed systems host marine methane hydrate
deposits. In such an ocean bed, sediments are exposed to a
pressure load and have experienced a certain history of this load.
These two factors result in compaction of the porous sediments. A
typical sample of such a compaction, increasing with depth, can be
seen in Fig.\,\ref{fig1}a, where the sediment porosity ($\phi$)
for different depths below the water-sediment interface is
reported~\cite{Proc_ODP-V164-InitResults,Proc_ODP-V164}. According
to~\cite{Davie-Buffett-2001,Davie-Buffett-2003} one can fit the
observed porosity with the exponential function
\begin{equation}
\phi(z)=\phi_0\exp(-z/L),
\label{eq-in-01}
\end{equation}
where $z$ is the depth below the water-sediment interface, $L$ is
the characteristic depth of compaction (see Caption to
Fig.\,\ref{fig1}).

Earlier approaches to the relation between porosity and transport
properties (such as, e.g., the Kozeny--Carman relation for
flow~\cite{Carman-1961,Mavko-Mukerji-Dvorkin-2009}) adopted
additional (though reasonable) constrains on the geometry of pore
channels and utilized several characteristic parameters (for
review see~\cite{Mavko-Mukerji-Dvorkin-2009}). As well as the
family of empirical equations relating permeability to porosity
and irreversible water saturation~\cite{Schlumberger-1991}, these
models do not provide information on how the characteristic
parameters will be transformed under compaction. Thus, a
particular theory of the latter is remaining required.

In this Letter, in order to reconstruct the relations between
parameters, we first argue for geometric features of compaction
and corresponding scaling laws. Then we derive the dependence of
transport coefficients on the porosity and, for a particular
example, apply the derived dependence to examine whether the
assumption of a linear temperature profile in sediments is
adequate. We find that the amended predicted rate of the
production of methane hydrate differs significantly from the
``traditional'' predictions assuming the linear temperature
profile; this finding is of importance for the assessment of the
planetary inventory of methane hydrate.

\section{Physical model and scaling laws: Compaction of porous medium}
Prior to constructing a mathematical description of the problem,
we need to establish physical features of the sediment compaction
in Fig.\,\ref{fig1}a.

In general, when considering the variation of transport
coefficients with porosity, one should keep in mind two critical
thresholds for porosity~\cite{Mavko-Mukerji-Dvorkin-2009}. First,
there is a critical porosity important for acoustic processes in
porous materials; a more porous material can be treated as a
``suspension'' and a less porous one is ``solid''. This first
critical porosity is material-specific and is typically near
$0.5$, {\it i.e.}, might be close to the lower edge of the part of
the sediment column in Fig.\,\ref{fig1}. For porosities lower than
the second threshold, chemical compaction is taking over the
mechanical one. This second transition occurs for porosities which
are beyond the scope of our study. According to seismic data, for
the part of the sediment column we consider, the first critical
transition does not occur as well. Indeed, geological and
mechanical structure of sediments is reported to be continuous
enough for using the seismic wave reflection to detect as small
volumetric fraction of bubbles in pores as
1--2\%~\cite{MacKay_etal-1994,Hovland_etal-1999,Proc_ODP-V164}.
Hence, qualitative transitions in the structure of porous
sediments are not expected.

The stress load on the system is anisotropic; the vertical
direction is discriminated due to gravity.  In a solid body the
stress would be strictly anisotropic.  However, the stress in
granular materials (including cemented ones) is known to be not
distributed homogeneously but to form force
chains~\cite{Liu-etal-1995}.  The stress along chains is much
bigger than the average stress, and the grain displacements and
deformations in the course of compaction are mainly driven by
these chains.  Meanwhile, the randomness of the geometry of
contacts and the branching of force chains decreases the
anisotropy of the network of force chains in comparison to the
stress anisotropy for a solid
body~\cite{Liu-etal-1995,Bouchaud-etal-2001,Genga-etal-2003}. For
simplicity, we assume the compaction process to be isotropic.
Henceforth, we treat an isotropic compaction and consider the
sediments to have similar geologic origin (which is reasonable for
the geological systems we consider, on the timescale of several
million years~\cite{Proc_ODP-V164}).

In our model, as a first approach to the problem, we do not
consider the fragmentation of sediment grains. Anisotropy of
compaction and fragmentation are the main reasons for the change
of the topology of the pore channels' network. Without them,
compaction affects this topology non-efficiently, but rather
shrinks the channels. Further, the variation of the density of the
solid matrix material is negligible against the background of the
pore shrinking.

Let us derive general scaling laws for the compaction of sediments
the physical features of which are described above. We consider a
certain volume $V$ of the porous medium and fix this volume to
particles of the solid matrix ($V$ changes due to compaction);
$L$, $l$, and $V_s$ stand for the total channels' length in $V$,
the characteristic channels' diameter, and the volume of the solid
material in $V$, respectively. Hence,
\\
(i) porosity
$$
\phi=\frac{L\,l^2}{V}=\frac{L\,l^2}{V_s+L\,l^2},
\quad\mbox{ and thus }\quad
L\,l^2=\frac{\phi V_s}{1-\phi}.
$$
(ii) Owing to the unchanging shape of the channels' network,
$L^3\propto V$.
\\
(iii) Owing to the unchanging density of the solid material,
$V_s=(1-\phi)V$ is constant in the course of compaction; therefore
$V\propto(1-\phi)^{-1}$ and
\begin{equation}
 L\propto\frac{1}{(1-\phi)^{1/3}}.
\label{eq-an-01}
\end{equation}
From (i),
\begin{equation}
l^2\propto\frac{\phi}{1-\phi}\frac{1}{L}\propto\frac{\phi}{(1-\phi)^{2/3}}.
\label{eq-an-02}
\end{equation}

These relations are obviously relevant for a tube model of the
porous medium. However, they are derived without actual relaying
on features of any specific model and we expect them to be
reasonably accurate for realistic geometric models of porous
medium~\cite{Roberts-Schwartz-1985} as long as we deal with a
sparse porous structure (in Fig.\,\ref{fig1} porosity varies in
the range from $0.5$--$0.7$). The scaling of the linear measure
$l$ for realistic models describes the variation of transversal
linear measures of pores. For low-porosity materials the actual
scaling laws become sensitive to features of the pore geometry
(e.g., see~\cite{Roberts-Schwartz-1985}).

\subsection{Permeability}
We use the definition of the permeability $K$
according to the Darcy's law ({\it e.g.},
see~\cite{Whitaker-1986,Nield-Bejan-2006}):
$$
\vec{v}_f=-(K/\eta)\nabla p,
$$
where $\vec{v}_f$ is the {\em filtration velocity} of the fluid,
$\eta$ is the dynamic viscosity, $p$ is pressure. When the
distribution of the orientation of pore channels and topology of
the pore network are unchanged---as they are in our model of
compaction---the fluid speed in pores $u_f\equiv\phi^{-1}v_f$
requires the pressure gradient proportional to $l^{-2}$. Thus,
$v_f\propto\phi l^2|\nabla p|$ and
\begin{equation}
K\propto\frac{\phi^2}{(1-\phi)^{2/3}}.
\label{eq-an-03}
\end{equation}
For instance, the data reported in Fig.\,\ref{fig1} yield
$K(\phi=0.69)/K(\phi=0.49)\approx 2.8$ due to compaction; that is
the permeability of the upper sediment zone ($\phi\approx 0.69$)
is by factor $2.8$ larger than that at the bottom of the shown
sediment column ($\phi\approx 0.49$).

Remarkably, in~\cite{Garg_etal-2008} a rough dependence of the
permeability on the porosity, $K\propto\phi^2$ [cf
Eq.\,(\ref{eq-an-03})], was adopted because of the lack of
reasonable theories on scaling laws for sediments experiencing
compaction.

\subsection{Molecular diffusion of solute}
The evolution of the solution
concentration $C$ in quiescent pore water is governed by the
equation
\begin{equation}
 \frac{\partial}{\partial t}(\phi C)
 =\nabla\cdot\left[\gamma_D(\phi)D_m\nabla C\right],
\label{eq-an-04}
\end{equation}
where $D_m$ is the molecular diffusivity of the solute in bulk,
and $\gamma_D(\phi)$ is the geometric factor featuring the pore
network.

Similarly to the permeability, the geometric factor for the
molecular diffusivity ($\gamma_D$) depends on porosity $\phi$
which varies significantly with depth. The length of the channels
$L$ does not effect the diffusional flux until the statistics of
channel orientations are changed. With the concentration gradient
given, the solute flux through the area $S$ is linearly
proportional to the area of channel cross-section,
$S_\mathrm{pore}$:
\begin{equation}
\gamma_D\propto S_\mathrm{pore}/S=\phi\,.
\label{eq-an-05}
\end{equation}
Recall, porosity $\phi$ is exactly the average value of
$S_\mathrm{pore}/S$. This becomes evident if one considers the
cubic volume thinly sliced parallel to one of its sides; for each
slice the fraction of the pore volume is $S_\mathrm{pore}/S$, and,
thus, for the whole cube volume the ratio of pore volume to the
cube volume, which is porosity $\phi$, is the average value of
$S_\mathrm{pore}/S$.

\subsection{Hydrodynamic dispersion}
Due to the irregularity of the
microstructure of the pore network, the macroscopically uniform
displacement of liquid results in a mixing flow in pores, which
acts as an additional diffusion and is referred to as hydrodynamic
dispersion~\cite{Saffman-1959,Sahimi-1993}. The hydrodynamic
dispersion in an isotropic medium is strictly anisotropic; the
longitudinal and lateral dispersion coefficients differ and are
linearly proportional to the filtration speed
$v_f$~\cite{Saffman-1959}:
$$
D_\parallel=v_fd_1,\qquad D_\perp=v_fd_2.
$$
For a steady viscous flow in pores, $D_{\parallel,\perp}\propto
u_f^2\tau_\mathrm{corr}\propto u_fl_\mathrm{corr}$ (recall, the
fluid speed in pores $u_f=v_f/\phi$). For an isotropic compaction,
$l_\mathrm{corr}$ is scaled as the pore network skeleton, that is
$\propto L$, and Eq.\,(\ref{eq-an-02}) yields
$D_{\parallel,\perp}\propto (v_f/\phi)L\propto
v_f/[\phi(1-\phi)^{1/3}]$. Hence,
\begin{equation}
d_{1,2}\propto\frac{1}{\phi\,(1-\phi)^{1/3}}.
\label{eq-an-06}
\end{equation}
Notice, the geometric factor $\gamma_D$ for molecular diffusion
[Eq.\,(\ref{eq-an-05})] and the hydrodynamic dispersion
coefficients [Eq.\,(\ref{eq-an-06})] are affected by the
compaction differently.

\subsection{Electrical conductivity}
For the electrical conductivity one should clearly distinguish two
cases: (i) sea water and (ii) pure water in pores. Due to
electrolytes dissolved, sea water possesses the electrical
conductivity about $5\rm{S/m}$ which is much more than that of the
porous-skeleton material. The current flows through the liquid
volume. On the pore scale, this case is geometrically equivalent
to the case of molecular diffusion. Indeed, for a steady diffusive
flux we have the equation $\Delta{C}=0$ for concentration $C$ in
the pore volume, zero normal derivative of $C$ on the pore walls,
and fixed mean (macroscopic) gradient of $C$; for electrostatic
potential $\varphi$ we find the same equation $\Delta\varphi=0$ in
the pore volume with zero normal derivative of $\varphi$ on the
pore walls and fixed macroscopic gradient of $\varphi$. For the
electrical conductivity $\sigma$ of {\em sea water}, this
equivalence yields
\begin{equation}
\sigma\propto\phi\,,
\label{eq-an-07}
\end{equation}
the same scaling law as Eq.\,(\ref{eq-an-05}). This law can be
observed for $\phi\gtrsim 0.15$ in realistic models of the pore
morphology and experiments (see Ref.\,\cite{Roberts-Schwartz-1985}
and references therein for experimental data).

The case of pure water is significantly more subtle. Without
electrolytes dissociated, water has small number of charge
carriers; the mineral surface conductivity can make significant
contribution into the microscopic electrical conductivity. For
sands, experiments demonstrate nearly the same conductivity for
wet massif and the massif fully saturated with pure water, which
indicates that the electrical current flows along the
water-mineral interface~\cite{Revil-Glover-1998}. In this case,
resistivity is mainly contributed by the sharp sand grain
contacts; the geometry of these contacts is controlled by many
factors, including the stress~\cite{Hertz-1882}. This problem is
very complex and lies beyond the immediate applicability field of
our results. Thus, we emphasize that Eq.\,(\ref{eq-an-07}) is
valid for salt water only.

\subsection{Heat diffusion}
Heat transfer in the porous medium is
governed by the equation
\begin{eqnarray}
 \frac{\partial}{\partial t}
  [(1-\phi)\rho_sc_{P,s}+\phi\rho_fc_{P,f})T]\qquad\qquad&&
 \nonumber\\
 +\nabla\cdot[\vec{v}_f\rho_fc_{P,f}T]
 =\nabla\cdot\left[\kappa(\phi)\nabla T\right],&&
\label{eq-an-08}
\end{eqnarray}
where $\rho_s$, $\rho_f$ and $c_{P,s}$, $c_{P,f}$ are the
densities and the specific heat capacities of the solid matrix and
the fluid, respectively, $\kappa(\phi)$ is the heat conductivity
of the fluid-saturated medium.

An evaluation of the dependence of the macroscopic thermal
conductivity on the porosity of sediments experiencing compaction
is much more complicated than for the permeability and the
solution diffusivity, because heat flows through both the solid
matrix and the fluid in pores and the fluxes in these two
subsystems (with complex random geometry) should be found and
conjuncted at the interface. This issue requires a particularly
accurate study. Straightforward scaling rules cannot be derived
with our approach here and, instead, we relay on empirical
relation suggested in Ref.\,\cite{Chaudhary-Bhandari-1969} for
porous media saturated with a weakly heat conducting fluid (such
as air of water, which has 5--10 times smaller heat conductivity
than typical mineral materials).

Chaudhary and Bhandari~\cite{Chaudhary-Bhandari-1969} reported the
law
\begin{equation}
\kappa(\phi)=\kappa_s\left(\frac{\kappa_f}{\kappa_s}\right)^{1-n}
 \frac{[\phi(1-\phi)]^n}{\phi+(1-\phi)\kappa_f/\kappa_s},
\label{eq-an-08}
\end{equation}
where $\kappa_f$ and $\kappa_s$ are the conductivities of matrix
material and fluid in pores, respectively, and
$n=0.5(1-\ln\phi)/\ln[\phi(1-\phi)\kappa_s/\kappa_f]$, to be
satisfactory accurate for a broad variety of sediment-kind porous
materials. For our study of carbon-bearing sediments, the relevant
matrix material heat conductivity is $\kappa_s=2.93\,{\rm
J/(m\,s\,K)}$~\cite{Chaudhary-Bhandari-1969} (for water
$\kappa_f=0.58\,{\rm J/(m\,s\,K)}$ is well known). In particular,
for the porosity profile in Fig.\,\ref{fig1},
$\kappa(\phi=0.69)/\kappa(\phi=0.49)\approx0.66$, {\it i.e.}, the
temperature profile slope varies by factor $1.5$ for different
parts of the sediment column.

\begin{figure*}[!t]
\centerline{
 \includegraphics[width=0.31\textwidth]%
 {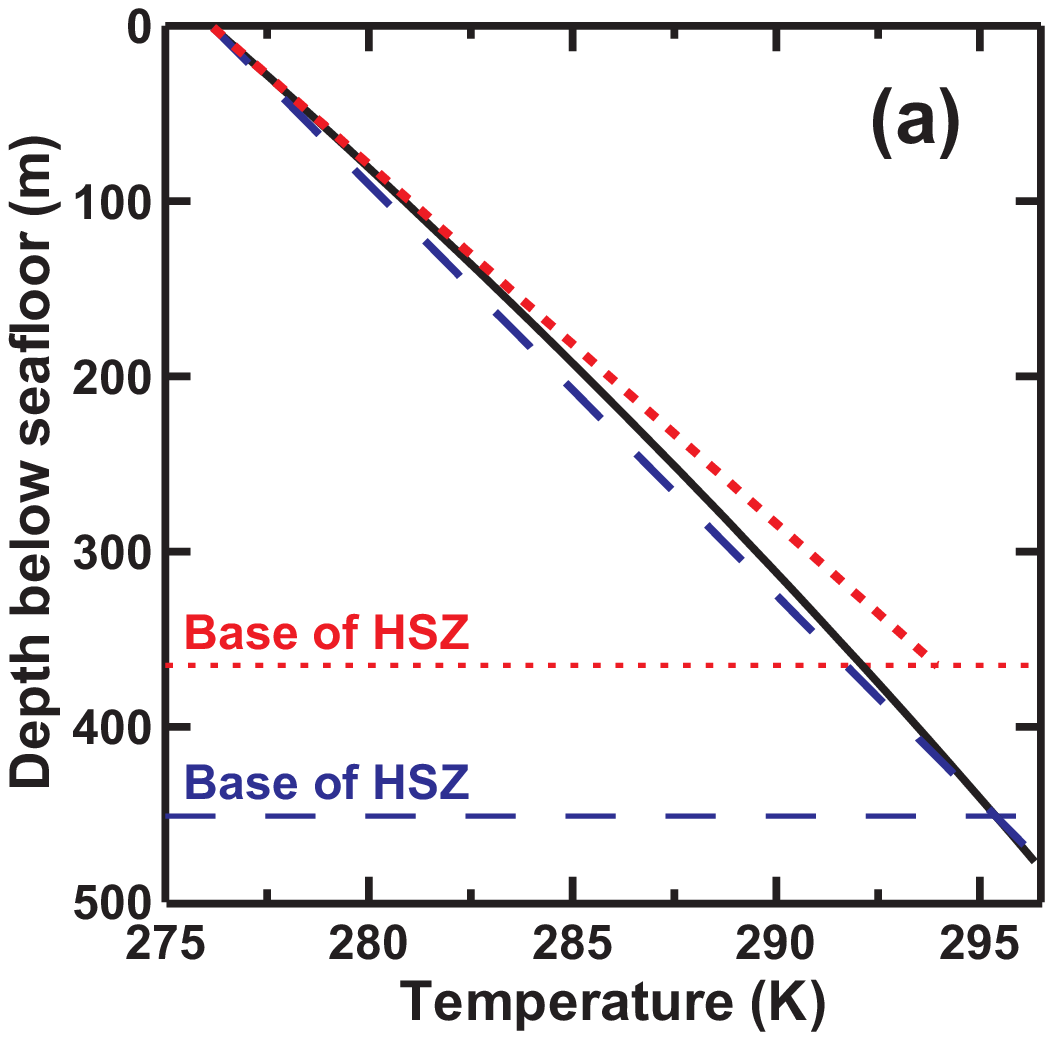}
\quad
 \includegraphics[width=0.31\textwidth]%
 {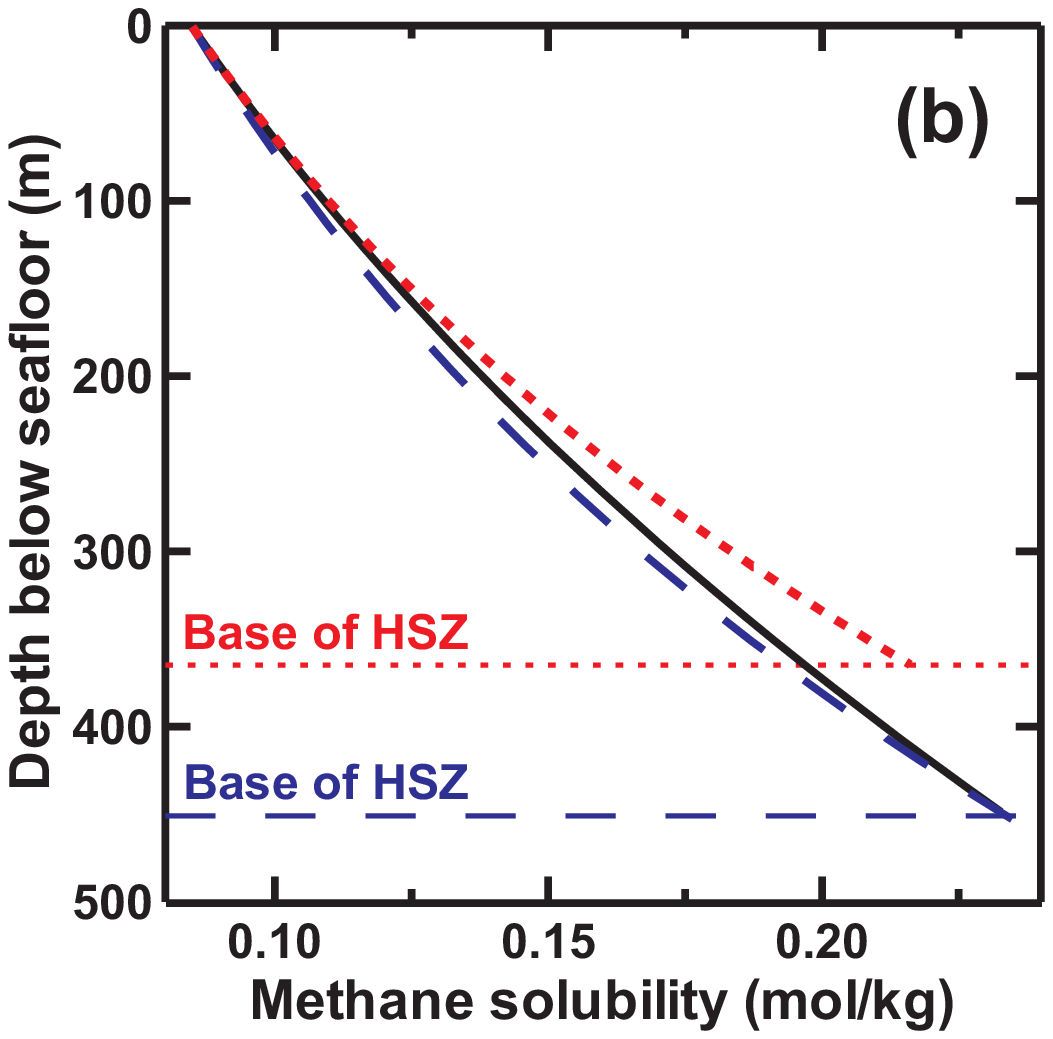}
\quad
 \includegraphics[width=0.31\textwidth]%
 {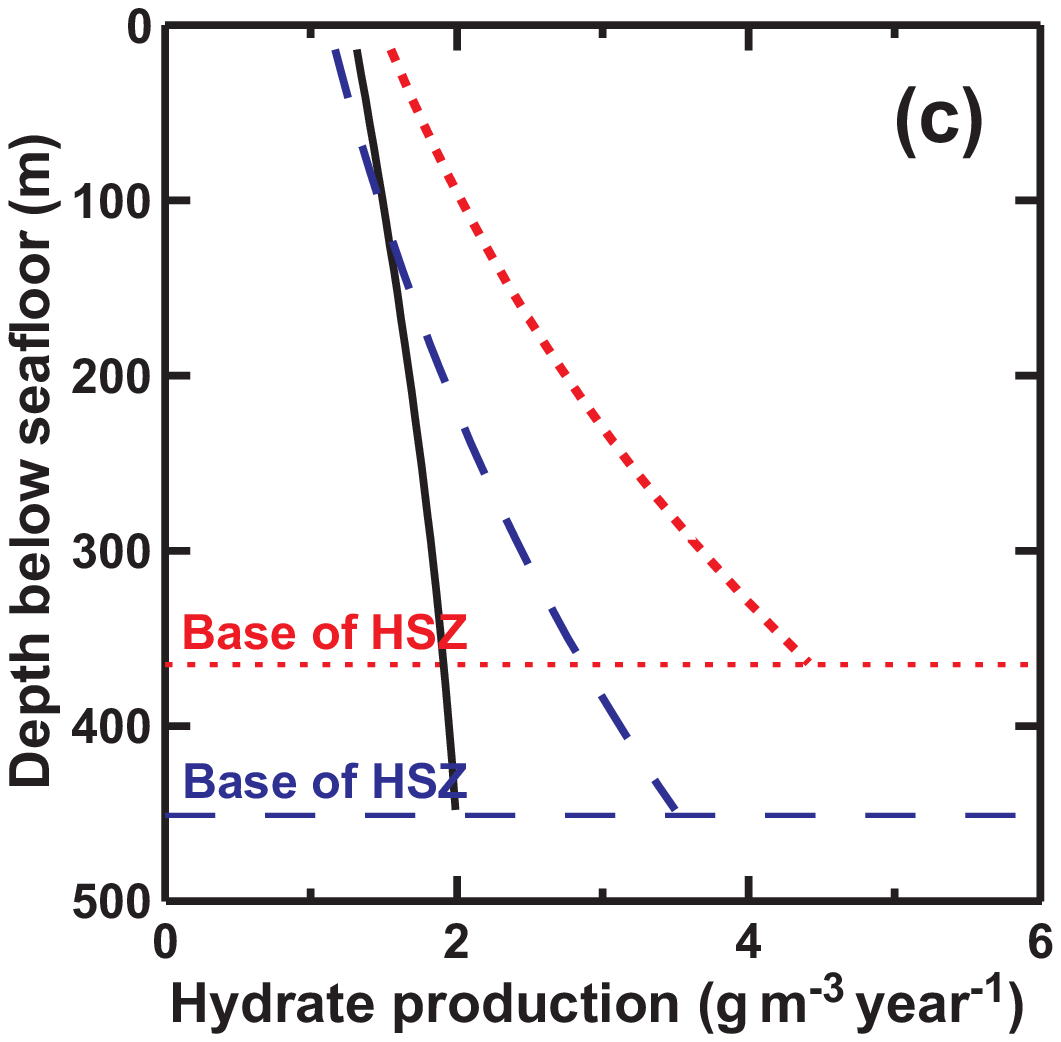}}

  \caption{(Color on-line)
Predictions for the data in Fig.\,\ref{fig1}: Temperature
profiles~(a), the aqueous methane solubility in equilibrium with
hydrate~(b), and the consequent production of methane hydrate due
to diffusion of methane dissolved in pore water~(c) for an
accurate nonuniform geothermal gradient (black solid lines), and
for simplified linear temperature profiles guessed either from the
observed position of the base of HSZ (blue dashed lines) or from
the observed geothermal gradient immediately next to the
water-sediment interface (red dotted lines).
 }
  \label{fig2}
\end{figure*}

\section{Geothermal gradient and methane hydrate inventory}
While the researchers modelling methane hydrate deposits ({\it
e.g.},
\cite{Davie-Buffett-2001,Davie-Buffett-2003,Archer-2007,Garg_etal-2008})
adopt a uniform geothermal gradient $G:=dT/dz$, it is, in fact,
nonuniform.  Instead, the heat flux, which is the product of the
geothermal gradient and the (nonuniform) thermal conductivity, is
uniform under time-independent conditions. Hence,
\begin{equation}
G(z)\equiv\frac{dT(z)}{dz}=\frac{\mbox{[heat flux]}}{\kappa(z)}
 =\frac{\kappa(0)}{\kappa(z)}\frac{dT(0)}{dz}.
\label{eq-im-01}
\end{equation}
Eq.\,(\ref{eq-im-01}) yields
\begin{equation}
T(z)=T(0)+G_0\int_0^z\frac{\kappa(0)}{\kappa(z_1)}dz_1,
\label{eq-im-02}
\end{equation}
where $G_0$ is the geothermal gradient next to the water-sediment
interface. The latter expression is convenient when $G_0$ is
directly measured in shallow upper layer of sediments. However,
for sediments bearing methane hydrate, the geothermal gradient is
practically derived in a different way.

Immediately below the floor of deep seas the pressure of the water
column is large enough and the temperature is low enough for
methane hydrate to be stable. Deeper into sediments the
temperature grows and, at certain depth, the pressure becomes not
sufficiently large to stabilize the hydrate (the critical pressure
depends on temperature nearly exponentially~\cite{Moridis-2003}).
Thus, the Hydrate Stability Zone (HSZ) spreads in sediments from
the water-sediment interface down to a certain depth, the base of
the HSZ, which is detected by the reflection of seismic
waves~\cite{MacKay_etal-1994}. Practically, for marine methane
hydrates the geothermal gradient is inferred from the position
$z=L_h$ of the base of the HSZ
\cite{Davie-Buffett-2001,Davie-Buffett-2003}:
$G:=(T(L_h)-T(0))/L_h$, where $T(L_h)$ is calculated as the
hydrate destabilization temperature for the known water salinity
and hydrostatic pressure $P(L_h)$ of the water column ({\it e.g.},
one can employ an accurate mathematical model of hydrate
from~\cite{Sun-Duan-2007}). For this case,
\begin{equation}
T(z)=T(0)+\big(T(L_h)-T(0)\big)\frac{\int_0^z\kappa^{-1}(z_1)\,dz_1}
 {\int_0^{L_h}\kappa^{-1}(z_1)\,dz_1}.
\label{eq-im-03}
\end{equation}

In Fig.\,\ref{fig2}a, the temperature profile consistent with
compaction is compared to ``traditional'' linear profiles guessed
either from the observed position of the base of HSZ or from
measurements of the sediments temperature profile next to the
water-sediment interface. Remarkably, assuming the geothermal
gradient being uniform is especially inaccurate for the latter
case (red dotted line): the assumption of the linear profile
significantly rises the base of the HSZ.

The role of the nonuniformity of $G$ appears to be especially
pronounced in the problem of hydrate formation. When a hydrate is
present in the HSZ, (i) the concentration of methane dissolved in
water is determined by the thermodynamic equilibrium between the
hydrate and the aqueous solution, in other words, it equals the
solubility, and (ii) the diffusive flux of the aqueous methane may
posses non-zero divergence equal to the formation/dissociation
rate of hydrate in pore water (up to a known constant multiplier
determining the fraction of methane in hydrate). We employed the
thermodynamic model of a hydrate developed in~\cite{Sun-Duan-2007}
for the calculation of the solubility profiles (Fig.\,\ref{fig2}b)
and then derived the contribution of this flux into the hydrate
production (Fig.\,\ref{fig2}c). In addition, to calculate the
divergence of the diffusive flux we accounted for the strong
dependence of the methane molecular diffusivity on the temperature
(the dependence on pressure is negligible):
 $D_m\approx(7+0.4\mathrm{K}^{-1}(T-273.15\mathrm{K})
 +10^{-3}\mathrm{K}^{-2}(T-273.15\mathrm{K})^2)\cdot10^{-10}
 \mathrm{m}^2/\mathrm{s}$,
which fits well with experimental data~({\it
e.g.},~\cite{Sachs-1998}). One can see, that linear temperature
profiles significantly overestimate the production of methane
hydrate in the lower part of the HSZ. Thus, the results of
mathematical modelling which simultaneously considers the
compaction of the sediment and ignores the consequent
non-linearity of the temperature profile ({\it e.g.},
\cite{Davie-Buffett-2001,Davie-Buffett-2003,Archer-2007,Garg_etal-2008})
are significantly affected by this inconsistency.

Notice, here we argue for the importance of the scaling laws for
compaction and assess the physical accuracy of models adopted
in~\cite{Davie-Buffett-2001,Davie-Buffett-2003,Archer-2007,Garg_etal-2008}
in this context only. This is the reason, why we readdress the
pure Fickian diffusion of aqueous methane and do not consider
non-Fickian effects, thermodiffusion and gravitational
stratification of the solute, ignored in the existing models of
the formation of hydrate deposits, although the importance of
non-Fickian effects was shown
in~\cite{Goldobin-Brilliantov-2011,Goldobin_etal-2011}.

\section{Conclusion}
Summarizing, we have described an isotropic compaction of porous
medium and derived scaling laws for geometrical properties of the
pore structure. These laws have yielded dependencies of transport
properties [permeability, Eq.\,(\ref{eq-an-03}), effective
molecular diffusivity, Eq.\,(\ref{eq-an-05}), hydrodynamic
dispersion, Eq.\,(\ref{eq-an-06}), electrical conductivity,
Eq.\,(\ref{eq-an-07}), and thermal conductivity,
Eq.\,(\ref{eq-an-08})] on the porosity for porous sediments of
similar geological origin. Notably, the compaction of sediments
(for example, see Fig.\,\ref{fig1}a) is an inherent feature of
most geological systems on the field scale. In particular, for
paradigmatic models of formation of marine methane
hydrate~\cite{Davie-Buffett-2001,Davie-Buffett-2003,Archer-2007,Garg_etal-2008},
compaction is a ``key ingredient''. The employment of our results
for transport coefficients provides an opportunity for a
significant enhancement of physical soundness and relevance of the
modelling of sediments experiencing compaction and, in particular,
the global methane hydrate inventory ({\it e.g.},
Fig.\,\ref{fig2}c demonstrates the inaccuracy of the hydrate
production rate in treatments disregarding the variation of the
thermal conductivity due to compaction).

\acknowledgements{
I thank N.\ V.\ Brilliantov, L.\ S.\ Klimenko, D.\ Packwood, J.\
Levesley, J.\ Rees, and P.\ Jackson for fruitful discussions and
comments. The work has been supported by NERC Grant no.\
NE/F021941/1.
 }

\bibliographystyle{eplbib}


\end{document}